\documentstyle[preprint,aps,epsfig]{revtex}
\def\beq{\begin{equation}}
\def\eeq{\end{equation}}
\def\beqa{\begin{eqnarray}}
\def\eeqa{\end{eqnarray}}

\def\GeV{\nobreak\,\mbox{GeV}}

\def\La{\Lambda}
\def\ka{\kappa}

\begin{document}
\draft
\title{\sc QCD Sum Rules Calculation of Heavy $\Lambda$ Semileptonic Decay}
\author { H.G. Dosch$^1$, E. Ferreira$^2$, 
M. Nielsen$^3$ and R. Rosenfeld$^4$}
\address{  $^1$Institut f\"ur Theoretische Physik, Universit\"at Heidelberg\\
 Philosophenweg 16, D-6900 Heidelberg, Germany\\[0.1cm]
 $^2$Instituto de F\'{\i}sica, Universidade Federal do Rio de  
Janeiro \\
C.P. 68528, Rio de Janeiro 21945-970, RJ, Brazil\\
 [0.1cm]
 $^3$Instituto de F\'{\i}sica, Universidade de S\~{a}o Paulo\\
  C.P. 66318,  05315-970 S\~{a}o Paulo, SP, Brazil \\[0.1cm]
 $^4$Instituto de F\'{\i}sica Te\'orica, Universidade Estadual Paulista \\
 Rua Pamplona, 145; 01405-900 -- S\~ao Paulo,SP, Brazil } 

\maketitle
 \begin{abstract}
We set up sum rules for heavy lambda decays in a full QCD calculation 
which in the heavy quark mass limit incorporates the symmetries of 
heavy quark effective theory. 
For the semileptonic 
$\La_c$ decay we obtain a reasonable agreement with experiment. 
For the $\La_b$ semileptonic decay  
 we find at the zero recoil point a violation of the heavy 
quark symmetry of about 20\%. 
\end{abstract}  
\bigskip
PACS Numbers~ :~ 12.38.Lg , 13.30.Ce , 14.65.Dw , 14.65.Fy
\newpage

The semileptonic decays of hadrons are the most valuable source of 
information for the determination of the Cabbibo-Kobayashi-Maskawa (CKM) 
matrix elements, but  
unfortunately these fundamental quantities of the standard model have 
to be disentangled from the effects of strong interactions occurring 
inside the hadrons. In the limit where the initial and final quarks are 
infinitely 
heavy the degrees of freedom can be integrated out and the remaining 
heavy quark effective theory (HQET) (\cite{VolSh87,NusW87,iw90,g90}, for a 
review see \cite{n94}) can make a number of rigorous 
statements  that are independent of the details of the strong interaction.
However, the corrections to the heavy quark effective theory due 
to finite quark masses are by no means negligible and therefore one has 
to look for more realistic treatments of the effects of strong interactions 
on the semileptonic decays. QCD sum rules \cite{SVZ79} (for reviews see 
\cite{Nar89,Shif92}) are one of the most widely used and best founded 
approaches for these purposes, treating  non-perturbative effects 
analytically with a limited 
input of phenomenological parameters, with good  success 
in the calculation of corrections to HQET
(see \cite{n94} and references therein). 
Particularly, QCD sum rules have shown that the non-leading corrections to 
HQET can be quite appreciable, even for hadrons with $b$-quarks 
\cite{bbbd92,Nar92}. 

Semileptonic decays of charmed into strange hadrons are among the best 
investigated processes of this kind. The relevant CKM matrix element 
$V_{cs}$ is well determined, so that 
calculations of these decays may provide very good tests  of the applied 
method. On the other hand, there are quite serious discrepancies between 
experiments and HQET in the ratios of the lifetimes of beautiful 
baryons to beautiful mesons 
(see e.g. \cite{neu97}), and hence it is of prime interest to investigate 
all decay channels of the $\La_b$ baryon.

In this letter we evaluate the semileptonic decays of heavy $\La$-baryons in 
the QCD sum rule approach as was developed for heavy meson decays in reference 
\cite{BBD91}. This approach treats full QCD but also reproduces the symmetries 
of HQET.  It is therefore a very interesting 
method for  the investigation of deviations from HQET in the decay 
$\La_b\to \La_c \ell \nu_{\ell}$.

We label generically the initial channel by $I$ and the final channel by $F$. 
For the decay  $\Lambda_I \to \Lambda_F + \ell + \nu_\ell$ we start from the 
three-point function of the weak transition  current from a initial to a 
final quark  $J_\mu= \bar Q_F \gamma_\mu (1- \gamma_5) Q_I$ and the 
interpolating fields of the initial and final baryons $\eta_{\Lambda_I}$ 
and $\eta_{\Lambda_F}$
\begin{equation}
\Pi _\mu (p_F,p_I)=(i)^2\int d^4xd^4y \langle 0|T\{\eta _{\Lambda_F}(x)
J_\mu (0)\overline{\eta }_{\Lambda _I}(y)\}|0\rangle 
e^{ip_Fx}e^{-ip_Iy}\; . 
\label{cor}
\end{equation}
The experimental information is contained in the decay amplitude \cite{Kor94}
\begin{eqnarray}
\langle\Lambda_F(p_F)|J_\mu|\Lambda_I(p_I)\rangle&=& \nonumber\\
&&\hspace{-3cm}\overline{u}(p_F)
[\gamma_\mu
(F_1^V+F_1^A\gamma_5)+i\sigma_{\mu\nu}q^\nu(F_2^V+F_2^A\gamma_5)+
q_\mu (F_3^V+F_3^A\gamma_5)]u(p_I)\; ,
\label{ff} 
\end{eqnarray}
where $q=p_F-p_I$ and the form factors are functions of $t=q^2$ .

We relate the theoretical (Eq. (\ref{cor})) with the physical 
(Eq. (\ref{ff})) 
quantities by inserting intermediate states into Eq. (\ref{cor}), 
which is then evaluated  in the not so deep Euclidean region where 
$p_F^2<M_{\Lambda_F}^2$ and $p_I^2<M_{\Lambda_I}^2$.

We introduce the couplings $f_{F_1}$ and $f_{I_1}$  of the currents with the 
respective hadronic states 
\begin{eqnarray}
\langle 0|\eta _{\Lambda_F}|\Lambda(p_F)\rangle &=&f_{F_1} u(p_F) 
\label{lam}
\\
\langle\Lambda_I(p_I)|\overline\eta _{\Lambda_I}|0\rangle &=& f_{I_1}
\overline{u}(p_I)\; ,
\label{lamc}
\end{eqnarray}
and obtain the phenomenological representation of Eq. (\ref{cor})
\begin{eqnarray}
\Pi_\mu^{(\rm phen)}(p_F,p_I) &=&{(f_{F_1}\rlap{/}{p_F}+f_{F_2})
\over p_F^2-
M_{\Lambda_F}^2}\left[\gamma_\mu
(F_1^V+F_1^A\gamma_5)+i\sigma_{\mu\nu}q^\nu(F_2^V+F_2^A\gamma_5)\right.
\nonumber \\*[7.2pt]
&+ & \left.
q_\mu (F_3^V+F_3^A\gamma_5)\right]{(f_{I_1}\rlap{/}{p_I}+f_{I_2})
 \over p_I^2-M_{\Lambda_I}^2} + \mbox{higher resonances}\; ,
\label{ficor}
\end{eqnarray}
where we have defined $f_{F_2}=f_{F_1}M_{\Lambda_F}$ and $f_{I_2}=f_{I_1}
M_{\Lambda_I}$.

The theoretical expression is evaluated by performing the operator 
product expansion of the operator in Eq.~(\ref{cor}) and then taking 
the expectation value with respect to the physical vacuum. The term from 
the unit operator gives the usual perturbative contribution, while  the 
vacuum expectation values of the other operators in the expansion give 
the nonperturbative corrections proportional to the condensates of the 
respective operators. Thus 
\beq
\Pi^{\rm theor}_\mu = \Pi^{\rm pert}_\mu + 
      \sum_i \Pi^{{\rm nonpert}(i)}_\mu ~ ,
\label{theor1}
\eeq
where the index i refers to the dimensions of the condensates.

As usual we evaluate the form factors $F^{V,A}_a$ of Eq. (\ref{ff}) 
($a= 1,2,3$) by matching the phenomenological representation Eq. (\ref{ficor}) 
of the three point function with the theoretical counterpart in 
Eq. (\ref{theor1}).
 We project out sum rules for the products $F^{V,A}_a(q^2) f_{I_i}\,f_{F_k} 
\; \;(i,k = 1,2)$ of the invariant amplitudes in  
Eq.~(\ref{ff}) and the current couplings $f$ of Eqs.~(\ref{lam},\ref{lamc}), 
by performing appropriate traces of Eq.~(\ref{ficor}). 
We thus obtain four sum rules for
each amplitude $F^{V,A}_a(t)$, but we use only those based on 
$f_{I_1}$ and 
$f_{F_1}$ since for them the imaginary part is positive definite. After this 
projection has been performed the treatment follows very closely the lines   
given in ref. \cite{BBD91}.   
We make the usual assumption that the contributions of the higher resonances 
(and the continuum) can be adequately approximated by the perturbative 
contributions above certain thresholds $s = p_I^2 \geq s_0$ and 
$u= p_F^2 \geq u_0$ and we use the Borel improvement by performing a 
double Laplace transform of the spectral functions of the theoretical 
expressions. A crucial ingredient for the incorporation of the HQET 
symmetries is to express the current couplings  $f_{I_i}$ and 
$f_{F_k}$ also through  QCD sum rules and relate 
the Borel parameters in the same way as explained in reference 
\cite{BBD91}.  This also leads to a considerable reduction of the
sensitivity to input parameters, like the continuum thresholds $s_0$ and
$u_0$, and to radiative corrections \cite{BBG93}.

 As it is well known from two-point sum rules for baryons
\cite{Iof81,CDKS82,bcdn92},  there is a continuum
 of choices for the interpolating currents. Of course the results should in 
principle be independent of the choice of the current (except for 
pathological cases which couple very weakly to the ground state), but the 
justification of the approximations depends on the choice made. In this 
letter  we concentrate on a very simple interpolating current where the 
two light quarks form a singlet spin and isospin  state, namely 
\beq
\eta_{\La_Q} = \epsilon_{ABC}(\bar u^A \gamma^5 d^B) Q^C ~,
\label{curr}
\eeq
where $u^A$ and $d^B$ stands for the Dirac field of light quarks of colours 
 $A$ and $B$, and $Q^C$ represents  a heavy quark ($b$ or $c$) of color $C$. 
This current couples strongly to the $\Lambda$ states in the heavy quark 
limit.  With this choice of current the quark condensate and the mixed 
gluon-quark condensate do not contribute to Eq.(\ref{theor1}). The 
gluon condensate does 
contribute, but experience with baryonic two-point functions and mesonic 
three point functions teaches us that it is of little influence. We are thus 
left with only the perturbative and the four quark condensate contributions.

In order to estimate the four quark condensate we use the factorization
\beq
\langle \bar{d}^A_\alpha\bar{u}^B_\beta u^{B'}_{\beta'}d^{A'}_{\alpha'} 
\rangle = {\kappa\over12^2}\delta_{\beta\beta'}\delta_{\alpha\alpha'}
\delta^{AA'}\delta^{BB''} \langle\bar{u}u\rangle\langle\bar{d}d\rangle ~ ,
\label{cond}
\eeq
where the parameter $\kappa$, which ranges from $1$ to $3$, represents 
the deviation from the factorization hypothesis \cite{Nar89}. For the 
value of the quark condensate 
we take $  < q \bar{q}> = -(230 \; {\rm MeV})^3$.

As general results we obtain
\beq 
\label{general}
F^A_1(t) =-\, F^V_1(t); \; F^V_2(t)= F^V_3(t)= F^A_2(t)= F^A_3(t) = 0\; .
\eeq

In Fig. 1 we show  the behaviour of the contributions to the form factor  
$F_1^V$ at $t$=0 for the process $\La_c \rightarrow \La \ell \nu_{\ell}$
for $\kappa=1$  as function of the Borel mass $M_F^2$ .
We have used  $M_{\Lambda_c}=2.285$ GeV, $M_{\Lambda}=1.115$ GeV,
 $m_c=1.4$ GeV, $m_s=0.17$ GeV and $V_{cs} = 0.975$. 
We observe that the
contributions from the continuum and from the four quark condensate are 
comparable and tend to stabilize each other for values  
$M_F^2\geq 5\; \GeV^2$. This seems to be a rather large value for 
the Borel mass. However, the influence of the 6-dimensional 
condensate is still large at that value and we expect the contributions of 
higher dimensional condensates to be  very important at smaller Borel masses.
 A classical sum rule window where the perturbative and non-perturbative 
contributions are in equilibrium is thus the range above $5$  GeV$^2$. 
In that range we obtain $F^V_1(0) =0.51 \pm 0.02$ for $\ka=1$.
We have also calculated the $t$- dependence of this form factor in the range 
$0\leq t \leq 0.6 \; \GeV^2 $ where we   do not 
encounter difficulties with non-Landau singularities \cite{BBD91}. The 
$t$-dependence is represented with dots in Fig. 2 for two different choices 
of the four quark condensate ($\kappa=1, 2$ ). It can be very well 
approximated by a pole fit (solid line for $\ka=2$ and dashed line for 
$\ka=1$). Using this form factor we obtain for the width
\beq
\Gamma(\La_c\to \La + e^+ +  \nu_e ) = (1.0  \pm 0.3) \times 10^{-13} \;
{\rm GeV}. 
\eeq
Within the errors this value agrees with the reported experimental 
value\cite{PDG97}
\beq
\Gamma(\La_c\to \La + e^+ +  \nu_e ) = (0.74  \pm 0.15) \times 10^{-13} \; 
{\rm GeV}.
\eeq

The values in Eq. (\ref{general}) yield an asymmetry parameter $\alpha = -1$ 
whereas the observed value \cite{PDG97} is  $-0.82\pm 0.10$.

The overwhelming source of the theoretical error is the uncertainty in  the 
value of the four quark condensate. Variations of the quark masses and 
continuum thresholds within the reasonable limits $0.13\leq m_s \leq0.17\GeV$, 
$1.25\leq m_c\leq 1.45\GeV$, $7.8\leq s_0\leq8.9\GeV^2$ and $2.6\leq u_0\leq
3.4\GeV^2$, are negligible as
compared to the errors introduced by the variation of the four quark 
condensate.

Considering $\kappa$ in the range $1\leq\kappa\leq3$ the form factor at the 
zero recoil point $t_{\rm max}= (M_{\Lambda_c}-M_{\Lambda})^2$ is 
$F^V_1(t_{\rm max}) = 0.79\pm 0.07$ , and  is less dependent on the input 
parameters than for smaller values of $t$. 

The form factors and the decay width of the semileptonic $\La_b$ decay was 
 calculated in the same way. We have used $M_{\Lambda_b}=5.65$ GeV, 
$M_{\Lambda_c}=2.285$ GeV,
$m_b=4.6$ GeV, $m_c=1.4$ GeV and $V_{cb} = 0.04$.

The behaviour of the OPE contributions to the form factor  
$F_1^V$ at $t$=0 for the process $\La_b \rightarrow \La_c \ell \nu_{\ell}$
as function of the Borel mass $M_F^2$ is shown in Fig. 3 for $\kappa=1$. 
Here the  relative importance 
of the four quark condensate is smaller than in the $\La_c$ decay.
In this case the region in the Borel mass where the perturbative and 
non-perturbative contributions are in equilibrium is above $10$  GeV$^2$ 
and we obtain $F^V_1(0) = 0.36 \pm 0.02$ for $\kappa=1$.
The $t$-dependence of this form factor can be calculated in the range 
$0\leq t \leq 8 \; \GeV^2 $ (which covers the major part of the 
kinematically allowed  region  $0\leq t \leq 11.34$ GeV$^2$) without 
encountering difficulties with non-Landau singularities \cite{BBD91}. The 
$t$-dependence is again well  approximated by  pole fits, as can be seen
in Fig. 4. The extrapolation of the fits to the  
maximal momentum transfer value, $t_{\rm max}=(m_{\La_b}-
m_{\La_c})^2$,  yields $F_1^V = - F^A_1 = 0.74$ for $\kappa=1$ or $2$. 
This value is remarkably
stable against variations of the input parameters like $\kappa$, $s_0$,
$u_0$ and the Borel mass. In the interval $1\leq\kappa\leq3$
we estimate the errors to be
\beq
F_1^V(t_{\rm max})=-F^A_1(t_{\rm max})=0.76\pm0.05\; .
\eeq
In HQET this value is just the Isgur-Wise function at zero quark recoil
and is predicted to be $1$. We thus find
a strong deviation of the heavy quark symmetry in the 
semileptonic decay $\La_b \to \La_c$. The deviation is larger than in the
$B \to D$ decays, where the corresponding value is about $0.9$, in accordance 
with sum rules  in full QCD\cite{Bal92}, and corrections to HQET\cite{Neu94}.
We have checked that this strong deviation is almost entirely due to the 
small mass of the charmed quark .
We have also calculated the semileptonic decay rate, which turns out to 
be much more sensitive  to the input parameters than the value at the zero 
recoil amplitude. We have obtained for the decay width
\beq
\Gamma(\La_b\to \La_c+ e + \bar \nu_e ) = (1.8 \pm 0.3) \times 10^{-14} \; 
{\rm GeV} ~ ,
\eeq
where the errors reflect variations of $\kappa$ from 1 to 3 and of the other 
parameters within reasonable limits. This value is consideraby smaller 
than other predictions \cite{Kor94,ILKK97,DHHL96} , which range from 
$3.5$ to $6$ $\times 10^{-14}$ GeV . It is interesting 
to note that the $1/m_Q$ corrections to HQET tend to increase the 
width\cite{Kor94,DHHL96}, whereas our result clearly indicates that the  
width in full QCD is
smaller  than in HQET. The decay width obtained by us stays well below the 
experimental  upper limit\cite{PDG97} given by $\Gamma(\La_b\to \La_c^+ 
\ell^-\bar\nu_{\ell} + {\rm anything})= (5.8 \pm 0.2)\times 10^{-14} $ GeV.

Important violations of HQET for some form factors are not unplausible 
due to the large difference between the mass of the $\Lambda_b$ and 
the $b$-quark ( see ref. \cite{n94}, table 4.1) , but such a large deviation 
for Luke protected form factors like $F_1^A(t_{\rm max})$ is certainly not 
expected. Although we can check that our treatment repects HQET we cannot, in 
the approach adopted here, evaluate reliably the $1/m_Q$ and $1/m_Q^2$ 
corrections. We 
can only state that a full QCD calculation which respects HQET and gives 
reasonable results for the experimentally known semileptonic decay of the 
$\Lambda_c$ indicates a large violation of heavy quark effective symmetry.

A detailed description of our procedure and the application to other decays 
and observables, as well as the dependence of the results with the
interpolating fields,
will be given in a forthcoming publication\cite{DFNR98}.

{\bf Acknowledgements} We thank  L.A. Barreiro   
for discussions. Financial support from  CNPq, USP, FAPESP, FAPERJ (Brazil)
 and DAAD (Germany) is gratefully acknowledged.

\begin{figure} 
\label{fig1}
\caption{The sum rule values for the decay amplitude $F_1^V$ at $t$=0 
for the process $\La_c \rightarrow \La \ell \nu_{\ell}$
 as function of the Borel mass $M_F^2$ . The long-dashed line is the 
perturbative contribution, the short-dashed line that of the four quark 
condensate for $\kappa=1$ (see Eq. (\protect \ref{cond})). The solid 
line is the total contribution.}
\end{figure}

\begin{figure}
\label{fig2}
\caption{The decay amplitude $F_1^V$ for the process 
$\La_c \rightarrow \La \ell \nu_{\ell}$
 as function of the squared momentum transfer $t$ to the leptons.
Solid line: Pole fit $F^V_1(t) =6.872 /(10.01 -t)$ for $\kappa=2$ 
(see Eq.(\protect \ref{cond})) to the 
sum rule results (dots). Dashed line: the same for $\kappa=1$; 
$F^V_1(t) =2.393 /(4.716 -t)$.}
\end{figure}

\begin{figure} 
\label{fig3}
\caption{The sum rule values for the decay amplitude $F_1^V$ at $t$=0 
for the process $\La_b \rightarrow \La_c \ell \nu_{\ell}$ 
 as function of the Borel mass $M_F^2$ . The long-dashed line is the 
perturbative contribution, the short-dashed line that of the four quark 
condensate for $\kappa=1$ (see Eq. (\protect \ref{cond})). The solid 
line is the total contribution.}
\end{figure}

\begin{figure}
\label{fig4}
\caption{The decay amplitude $F_1^V$ for the process $\La_b \rightarrow 
\La_c \ell \nu_{\ell}$  as function of the squared momentum transfer $t$ 
to the leptons. Solid line: Pole fit $F^V_1(t) =15.32 /(32.03 -t)$ for
$\kappa=2$ to the  sum rule results (dots). Dashed line: the same 
for $\kappa=1$; $F^V_1(t) =8.12 /(22.27 -t)$.}
\end{figure}

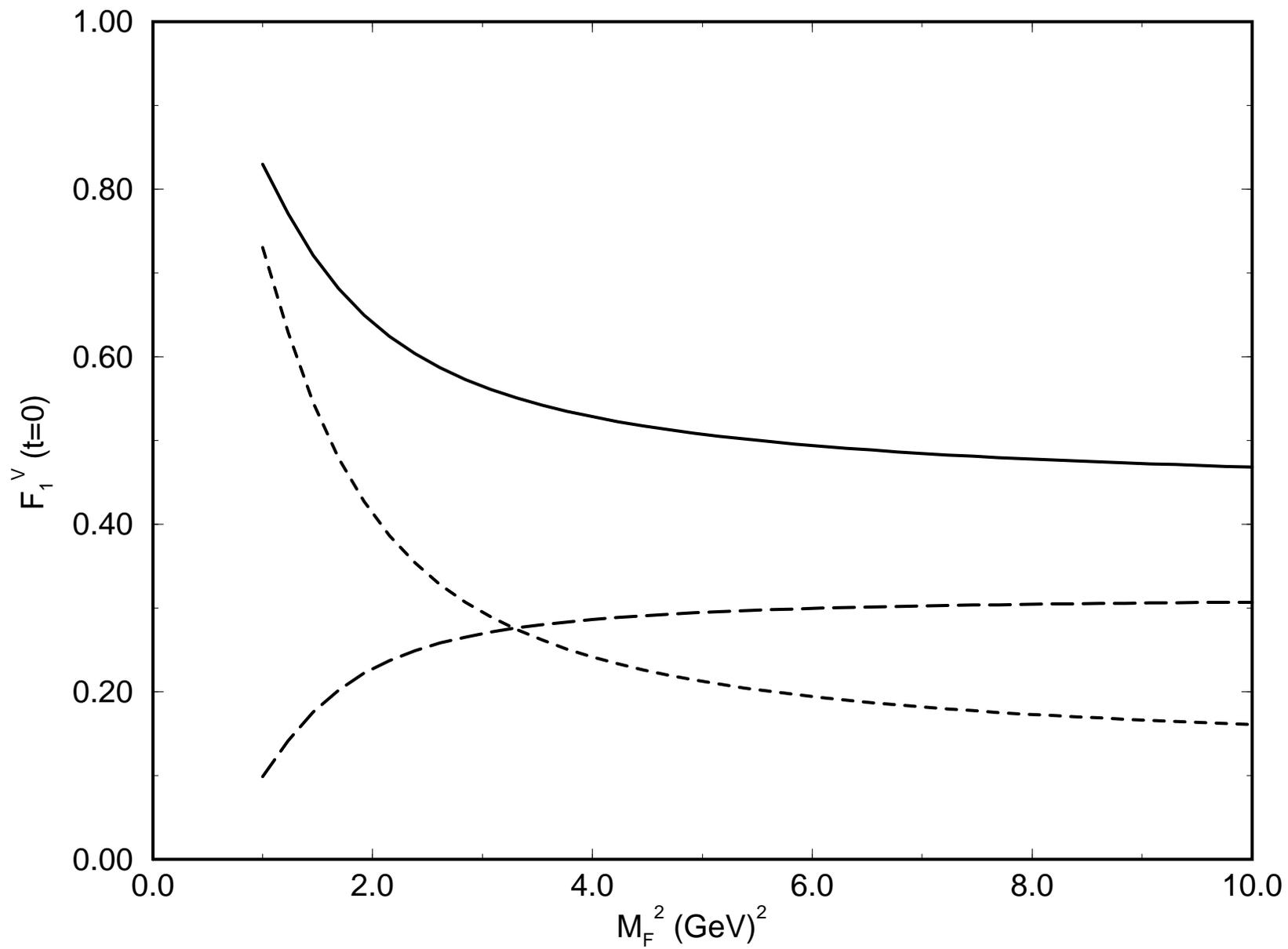
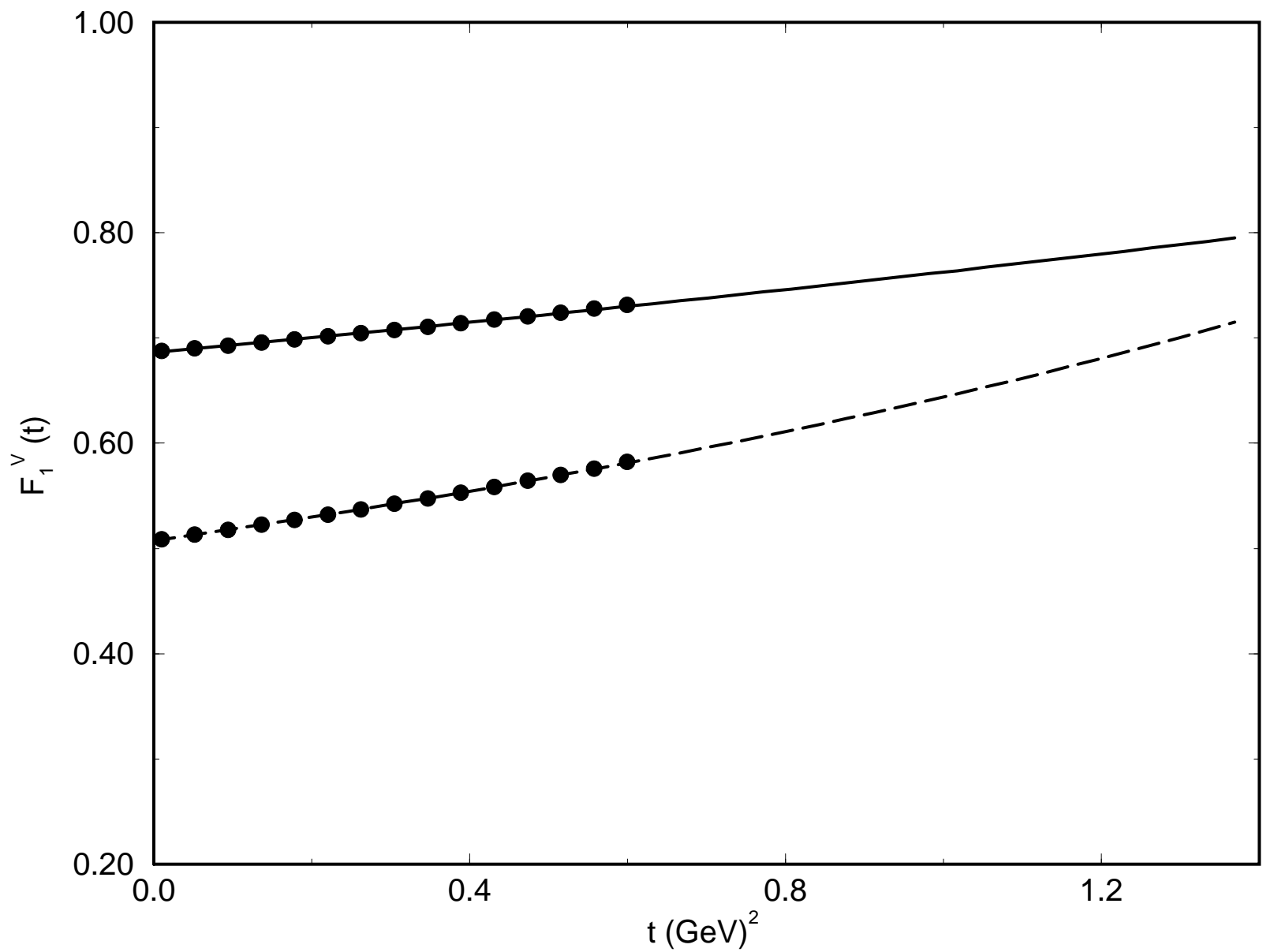
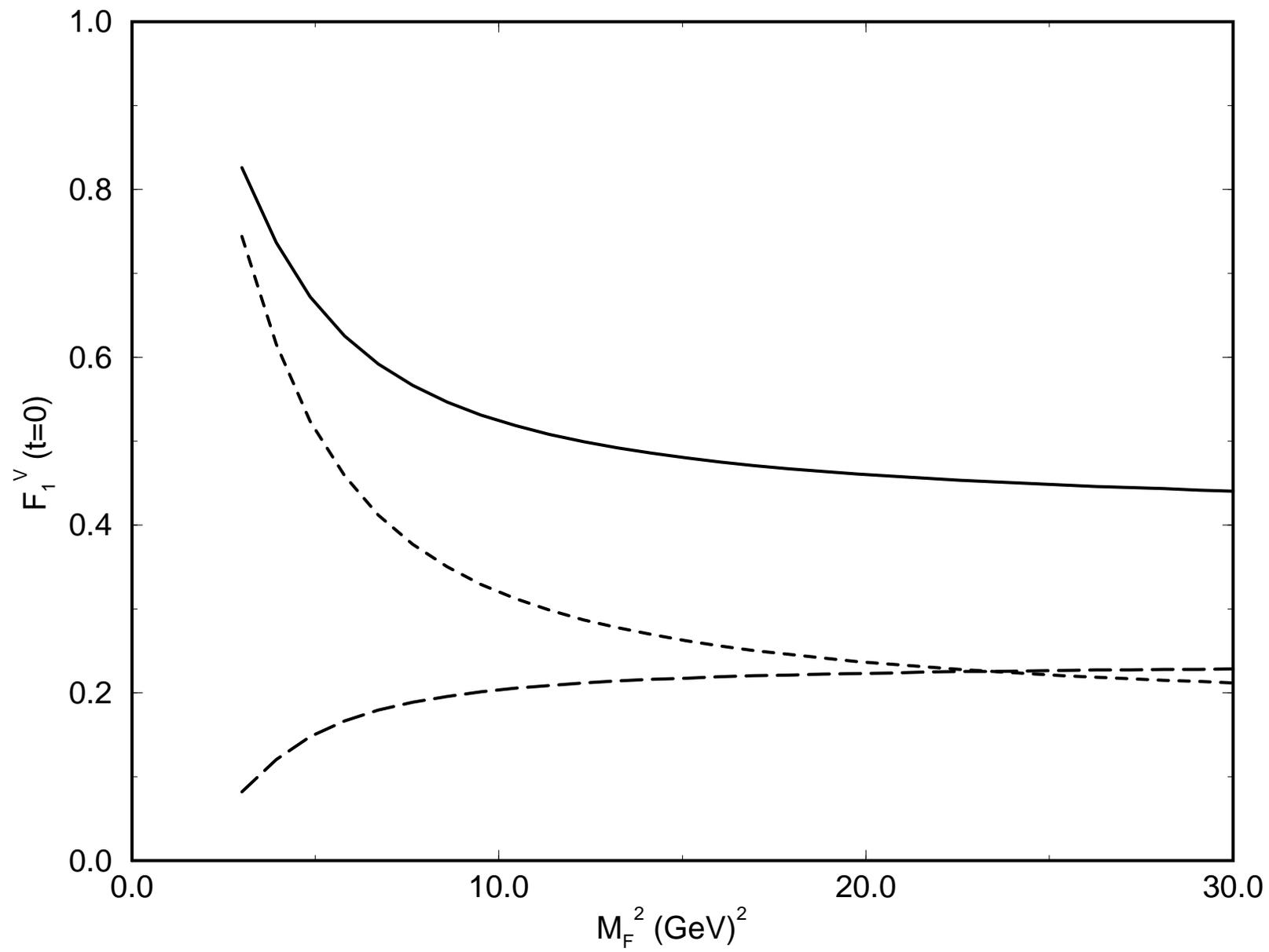
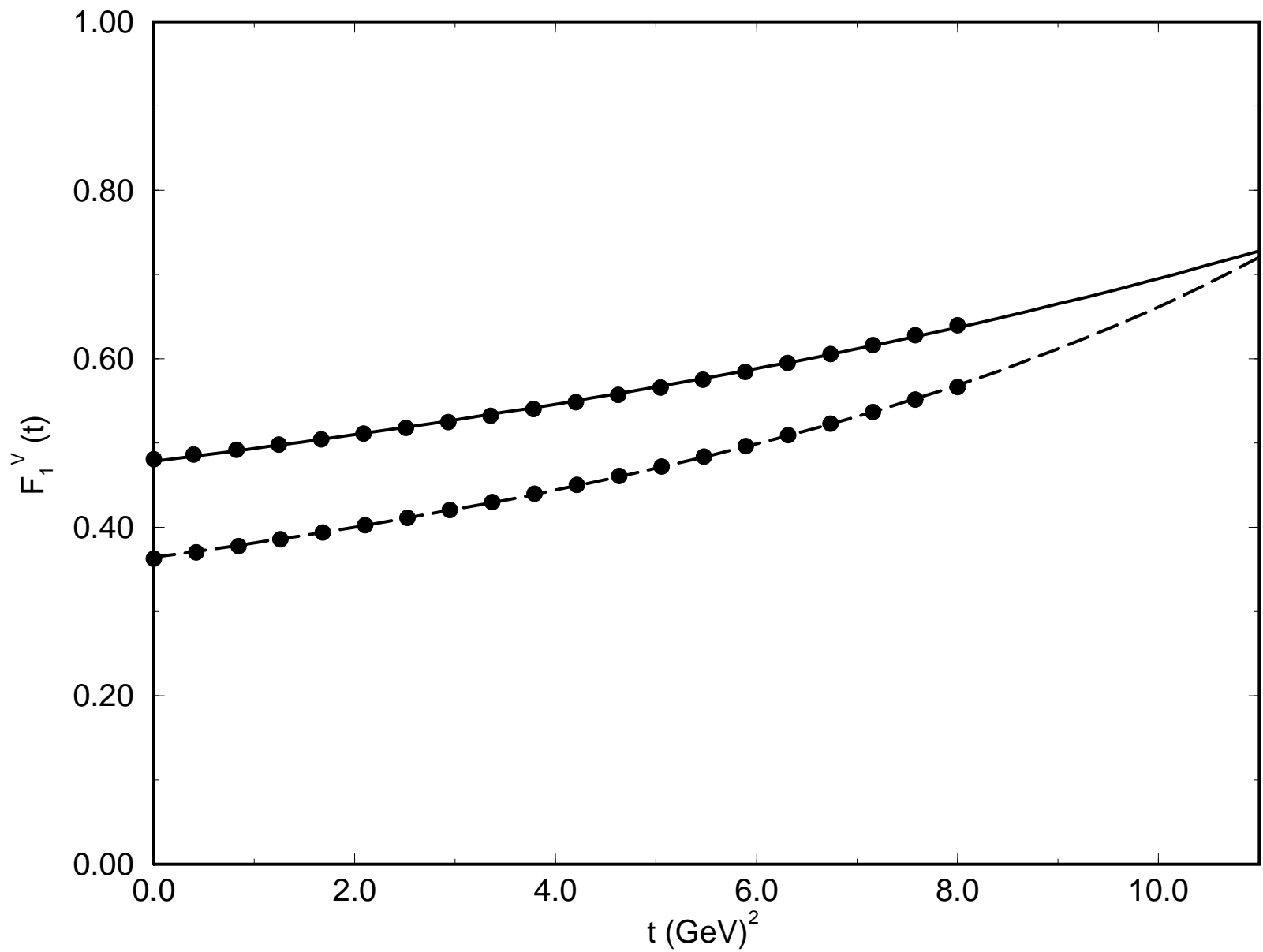
\end{document}